\documentclass[prb,preprint,showpacs,amssymb,floatfix]{revtex4}
\usepackage{epsfig}
\usepackage{graphicx}
\usepackage{dcolumn}
\usepackage{bm}
 
\begin{document}
\title{The Effect of Columnar Disorder on the Superconducting Transition of a Type-II Superconductor in Zero Applied Magnetic Field} 
\author{Anders Vestergren and Mats Wallin}
\affiliation{Condensed Matter Theory, Department of Physics, KTH,
SE-106 91 Stockholm, Sweden}
\author{S. Teitel}
\affiliation{Department of Physics and Astronomy, University of
Rochester, Rochester, NY 14627}
\author{Hans Weber}
\affiliation{Division of Physics, Lule{\aa} University of Technology, SE-971 87 Lule{\aa}, Sweden}
\date{\today} 
\begin{abstract}
We investigate the effect of random columnar disorder on the superconducting phase transition 
of a type-II superconductor in zero applied magnetic field using numerical simulations of three dimensional XY and vortex loop models.  We consider both an unscreened model, in which the bare magnetic penetration length is approximated as infinite, and a strongly screened model, in which the magnetic penetration length is of order the vortex core radius.  We consider both equilibrium and dynamic critical exponents.  
We show that, as in the disorder free case, the equilibrium transitions of the unscreened and strongly screened models lie in the same universality class, however scaling is now anisotropic.  
We find for the correlation length exponent $\nu=1.2\pm 0.1$, and for the
anisotropy exponent $\zeta=1.3\pm 0.1$.
We find different dynamic critical exponents for the unscreened and strongly screened models.
\end{abstract}
\pacs{}
\maketitle

\section{Introduction}

The discovery of high temperature superconductors, in which thermal fluctuations
are important and mean field theory can no longer be applied, has led to a resurgence
of interest in phase transitions and critical phenomena in type-II superconductors.\cite{Blatter}
Of particular interest has been understanding the effects of random quenched disorder
on the nature of the ordered phases and the universality of the phase transitions.  For
the high temperature superconductors, this quenched disorder can take many forms: random
point defects due to oxygen vacancies, planar twin grain boundaries, and columnar 
defects introduced by ion irradiation.  

Most of the work in this area has focused on
the case of a finite applied magnetic field, where one seeks to understand how the 
randomness distorts or destroys the Abrikosov lattice of magnetic field induced
vortex lines that forms in a pure system.  Columnar defects\cite{Nelson,Wallin1,Wallin2} 
have received considerable
attention, as they are particularly effective in pinning vortex lines and reducing flux
flow resistance.  In contrast, in this paper we will focus on the effect of columnar
defects on the superconducting phase transition in {\it zero} applied magnetic field.
We expect this case to be interesting for the following reason. A generalized Harris
criterion\cite{Wallin3,Chayes,Harris} argues that disorder will be a relevant perturbation, 
and change the universality
class of a phase transition, whenever $2-d^*\nu>0$, where $d^*$ is the number of
dimensions in which the system is disordered, and $\nu$ is the usual correlation
length critical exponent.  For a disorder free superconductor, the 
transition in zero applied magnetic field
is in the universality class of the three dimensional (3D) XY model,\cite{Dasgupta}
for which $\nu\agt 2/3$.  For random {\it point} disorder, $d^*=3$, so
$2-d^*\nu<0$, and the
generalized Harris criterion argues that the universality of the transition
remains unchanged.  For {\it columnar} disorder, however, $d^*=2$, and so
$2-d^*\nu>0$.  Columnar defects should therefore cross the zero field 
transition over to a new universality class.  Stability\cite{Chayes}
of this new disordered fixed point with respect to the generalized Harris criterion
implies that it should have a new value $\nu>1$.  In our equilibrium simulations
we indeed find behavior consistent with this, and we obtain a value for the
correlation length exponent $\nu=1.2\pm 0.1$.  Moreover, we find scaling
is now anistropic and we find the value of the anisotropy exponent to be
$\zeta=1.3\pm 0.1$.  Experimental measurement of these exponents would 
therefore provide a precision test of the theoretical model.

The model we study also has application to the $T=0$ superconductor to
insulator quantum phase transition in two dimensional thin films with random 
substrate disorder,\cite{Wallin3,Cha,Svistunov} and to the Mott
transition for bosons in 2D optical lattices with the addition of random scattered
laser intensity. \cite{optical}  In these cases, the two dimensional quantum
problem can be mapped onto a corresponding three dimensional classical
problem with the same symmetries as the one we study here.\cite{Sachdev}

To study the effect of columnar disorder on the zero field transition of a type-II superconductor,
we will consider two different limits.  The first is the limit of an ``unscreened"
superconductor\cite{Wallin2,Teitel}
in which magnetic field fluctuations are frozen out, corresponding 
to the approximation of an infinite bare magnetic penetration length, $\lambda_0\to\infty$. 
Here, vortex line segments have long ranged Coulombic-like interactions.
For the extreme type-II high temperature superconductors, for which $\kappa\equiv\lambda_0/\xi_0\sim 100$, where $\xi_0$ is the bare coherence length
that also sets the radius of a vortex line core,
the unscreened model should give a good description except 
in an {\it extremely} narrow temperature window about the transition.\cite{Fisher}

The second limit is that of a strongly screened superconductor,\cite{Wallin1,Wallin4}
corresponding to
the case $\lambda_0\sim\xi_0$.  In this case, vortex line segments have
short range interactions.  This description should also become valid
extremely close to the transition when the diverging correlation length $\xi$ becomes
comparable to the renormalized magnetic penetration length $\lambda$,
$\lambda\lesssim\xi$, and magnetic field fluctuations on such large length scales
must be included in 
determining the true critical behavior.  This region near $T_c$ may, however,
be too small to observe in practice.\cite{Fisher}

As in the disorder free case, a duality 
transformation\cite{Dasgupta,Kleinert,Savit} establishes 
that these two limits lie in the same universality class as regards equilibrium critical behavior.
They may be different, however, for dynamic critical behavior.\cite{Wallin4}  In this work we
carry out detailed Monte Carlo (MC) simulations of the XY model for the unscreened
superconductor to determine the equilibrium critical exponents, and we demonstrate
the presence of anisotropic scaling; by duality, these exponents also apply to the
strongly screened case.  Then, using simple local Monte Carlo dynamics, we compute
the dynamic critical exponent for both the unscreened XY model, and for the
strongly screened vortex loop model.  We find that the dynamic exponent is
different for these two limits.

The remainder of this paper is organzied as follows.  In Section II we describe
the XY model and the loop model for the unscreened and strongly screened limits, 
respectively.  The duality transformation between the two is given in Appendix A.
In Section III we discuss the equilibrium critical behavior of the XY model,
presenting our finite size scaling analysis, defining the observables we measure,
and giving the numerical results of our simulations.  In Section IV we discuss the
dynamic critical behavior of the XY and loop models, within a simple local Monte 
Carlo dynamics.  We
define the observables we measure and give our numerical results.  In Section V we give
our discussion and conclusions.

\section{Models}

\subsection{XY Model}
To model the effects of thermal fluctuations in a type-II superconductor, we start with the commonly used 3D XY model.\cite{Teitel}  This models the phase fluctuations of the superconducting order parameter in the ``unscreened" limit where magnetic field fluctuations are frozen out, corresponding to the approximation of an infinite bare magnetic penetration length, $\lambda_0\to\infty$.  For zero applied magnetic field we have,
\begin{equation}
{\cal H}_{\rm XY}[\theta_i]=-\sum_{i,\mu}J_{i\mu}\cos(\theta_i-\theta_{i+\hat\mu})\enspace.
\label{eHXY}
\end{equation}
Here $\theta_i$ represents the phase angle of the complex superconducting order parameter on node $i$ of a periodic cubic grid of $N=L\times L\times L_z$ sites, with periodic boundary conditions in all directions.  The sum is over all nearest neighbor bonds $(i,\mu)$ of the grid, 
with $\hat\mu=\hat x,\hat y,\hat z$, and the cosine term represents the kinetic energy of fluctuating supercurrents.  The short length cutoff of the discrete grid models the bare vortex core size $\xi_0$.  

In a pure system, the couplings $J_{i\mu}$ are all equal, except for a possible variation with bond direction $\mu$.  Here, we take the $J_{i\mu}$ randomly distributed in order to model quenched random columnar defects.  For the work reported on here, with columnar defects aligned parallel to the $\hat z$ axis, we have chosen the following distribution: in the $\hat z$ direction, we take all $J_{iz}=1$; in the $xy$ plane, we take $J_{i\mu}$, $\mu=x,y$, distributed equally likely with the two values $0.1$ and $1.9$, keeping the $J_{i\mu}$ translationally invariant along the $\hat z$ axis so as to model columnar disorder.  
Note that the random $J_{i\mu}$ introduce no frustration into the system; in the ground state all the $\theta_i$ are equal.
The variations in the $J_{i\mu}$ result in spatially random pinning energies for vortex loop excitations of the phase angles $\theta_i$.  We have chosen the above bimodal distribution
for $J_{i\mu}$ to give strong pinning energies (for fixed average $J_{i\mu}$),
so as to be able to approach the asymptotic scaling limit with reasonable size systems.  

Although we will simulate the Hamiltonian of Eq.(\ref{eHXY}) using periodic boundary
conditions on the phase angles $\theta_i$, it is useful to consider a more general
{\it fixed twist} boundary condition,
\begin{equation}
\theta_{i+L_\mu\hat\mu}=\theta_i+\Delta_\mu\enspace,
\label{eFTBC}
\end{equation}
where $\Delta_\mu$ is a fixed (non-fluctuating) total twist in the phase angle
applied across the system in direction $\hat\mu$.
Periodic boundary conditions correspond to the twist $\Delta_\mu=0$.
Transforming to new variables,
\begin{equation}
\theta^\prime_i=\theta_i-(\Delta_\mu/L_\mu){\bf r}_i\cdot\hat\mu\enspace,
\label{ethetap}
\end{equation}
the Hamiltonian of Eq.(\ref{eHXY}) becomes,
\begin{equation}
{\cal H}_{\rm XY}[\theta_i^\prime;\Delta_\mu]=-\sum_{i,\mu}J_{i\mu}\cos(\theta^\prime_i-\theta^\prime_{i+\hat\mu}-\Delta_\mu/L_\mu)\enspace,
\label{eHXY2}
\end{equation}
where the $\theta^\prime_i$ obey periodic boundary conditions,
$\theta^\prime_{i+L_\mu\hat\mu}=\theta^\prime_i$.
Using the fact that the cosine is periodic in $2\pi$, the partition function integrals over
$\theta_i^\prime$ can be taken over the interval $\theta_i^\prime\in[0,2\pi)$, as were
the integrals over the original phase angles $\theta_i$.  Considering how the free energy varies with the twist $\Delta_\mu$ will be useful later for discussing phase coherence in the model.

\subsection{Loop Model}

Although we carry out our equilibrium simulations directly in terms of the XY model of Eq.(\ref{eHXY}), we also consider a different formulation of the model.  If instead of the cosine interaction of Eq.(\ref{eHXY}), one uses the periodic Gaussian interaction of Villain,\cite{Villain} then a standard duality transformation\cite{Dasgupta,Kleinert,Savit} 
(see Appendix A) maps the XY model, ${\cal H}_{\rm XY}/T$, onto a 
model of sterically interacting loops, ${\cal H}_{\rm loop}/\tilde T$, where,
\begin{equation}
{\cal H}_{\rm loop} = {1\over 2}\sum_{i,\mu}g_{i\mu}n_{i\mu}^2\enspace.
\label{eHloop}
\end{equation}
The $n_{i\mu}$ are integer valued variables on the bonds $(i,\mu)$ and satisfy a divergenceless constraint,
\begin{equation}
\sum_{i,\mu}\left[n_{i\mu}-n_{i-\hat\mu,\mu}\right]=0\enspace.
\label{en}
\end{equation}
The $n_{i\mu}$ thus form connected paths through the system that must eventually close upon themselves. 
The couplings $g_{i\mu}$ of Eq.(\ref{eHloop}) are related to the couplings of the XY model by,
\begin{equation}
g_{i\mu}/\tilde T = T/J_{i\mu}\enspace,
\label{egi}
\end{equation}
where the temperature scale of the loop model, $\tilde T$, is inverted with respect
to the temperature scale of the XY model, $T$.


While the loop model of Eq.(\ref{eHloop}) is exactly dual to the 
XY model of  an {\it unscreened} superconductor,
taking it on its own with $\tilde T$ as the physical 
temperature, we can give ${\cal H}_{\rm loop}$  the following different physical interpretation.\cite{Wallin1,Wallin4}
We can regard the divergenceless variables $n_{i\mu}$ as the vortex loops of a
{\it strongly screened} superconductor with $\lambda_0\sim\xi_0$.  The short ranged
vortex line interaction of this case is then modeled by the simple onsite repulsion of ${\cal H}_{\rm loop}$.  Further details of this analogy may be found in Ref.\onlinecite{Wallin4}.
If we regard each site of our numerical grid as representing
a columnar pin, the random $g_{i\mu}$ in the $xy$ plane can be thought of as
modeling the random distances between such pins, and hence giving the random energies
associated with a vortex loop segment hopping from one pin to another.
This duality between ${\cal H}_{\rm loop}$ and ${\cal H}_{\rm XY}$ thus implies that the unscreened and the screened superconductor
models fall in the same equilibrium universality class, just as is the case for the disorder free model.\cite{Dasgupta}

\section{Equilibrium Critical Behavior}

In this section we report on our equilibrium XY model simulations.
To extract critical exponents, we use the method of finite size scaling. We first, therefore, discuss this method.

\subsection{Finite Size Scaling}

Because the columnar disorder singles out the special direction $\hat z$, we must allow for the possibility that scaling will be anisotropic.  If $\xi$ denotes the correlation length in the $xy$ plane, then anisotropic scaling assumes that, as $T\to T_c$ and $\xi$ diverges, the correlation length along the $\hat z$ axis diverges as,
\begin{equation}
\xi_z\sim\xi^\zeta\enspace,
\label{exiz}
\end{equation}
where $\zeta$ is the anisotropy exponent.

Consider now an observable ${\cal O}$ whose scaling dimension is zero.  As a function of reduced temperature $t\equiv (T-T_c)/T_c$ and system size $L\times L\times L_z$, we expect the scaling relationship,
\begin{equation}
{\cal O}(T,L,L_z)=\tilde{\cal O}(tb^{1/\nu},L/b,L_z/b^\zeta)\enspace,
\label{eScale}
\end{equation}
where $b$ is an arbitrary length rescaling factor, $\tilde{\cal O}$ is the scaling function, and $\nu$ is the usual correlation length
exponent,
\begin{equation}
\xi\sim t^{-\nu}\enspace.
\label{exi}
\end{equation}
Taking $b=L$ in Eq,(\ref{eScale}) then gives,
\begin{equation}
{\cal O}(T,L,L_z)=\tilde{\cal O}(tL^{1/\nu},1,L_z/L^\zeta)\enspace.
\label{eScale2}
\end{equation}
For the case of {\it isotropic} scaling, with $\zeta=1$, choosing a constant aspect ratio $L_z=\gamma L$ reduces the right hand side of Eq.(\ref{eScale2}) to a function of the single scaling variable $tL^{1/\nu}$.  Measuring ${\cal O}$ vs. $T$ for systems with varying $L$ but fixed $L_z/L$ is then sufficient to determine the exponent $\nu$.  However when $\zeta\ne 1$, and its value is unknown, it becomes necessary to consider systems with varying aspect ratio $L_z/L$, greatly increasing the complexity of the computation.  

To deal with this case we take the following approach, originally used to study
the phase transition in the quantum Ising spin glass.\cite{Young}  
Assume that the observable ${\cal O}(T,L,L_z)$ when viewed as a function of $L_z$, for fixed $T$ and $L$, has a maximum at a particular value $L_{z\, {\rm max}}$.  
Because of the scaling law Eq.(\ref{eScale2}), this value $L_{z \,{\rm max}}$ must occur when 
\begin{equation}
L_{z\,{\rm max}}/L^\zeta=\tilde\gamma(tL^{1/\nu})\enspace, 
\label{eLz}
\end{equation}
where $\tilde\gamma$ is a scaling function of the single variable $tL^{1/\nu}$.
We then define,
\begin{equation}
{\cal O}_{\rm max}(T,L)\equiv {\cal O}(T,L,L_{z \,{\rm max}})=\tilde{\cal O}(tL^{1/\nu},1,\tilde\gamma(tL^{1/\nu}))\equiv \tilde{\cal O}_{\rm max}(tL^{1/\nu})\enspace.
\label{eOmax}
\end{equation}
Plotting ${\cal O}_{\rm max}(T,L)$ vs. $T$ for different values of $L$, the curves will intersect at the common point $T=T_c$ (i.e. $t=0$).  The slopes of these curves at $T_c$ then
determine the exponent $\nu$.  In practice, we will determine the values of
$T_c$ and the exponent $\nu$ by the following approach.\cite{Nightingale}  Close to $T_c$ (i.e for
small $t$) we can
expand the scaling function $\tilde {\cal O}_{\rm max}$ 
as a polynomial for small values of its argument $tL^{1/\nu}$,
\begin{equation}
\tilde {\cal O}_{\rm max}(tL^{1/\nu})\simeq a_0 + a_1[(T-T_c)/T_c]L^{1/\nu}+a_2\left([(T-T_c)/T_c]L^{1/\nu}
\right)^2+\dots
\label{epoly}
\end{equation}
We then  fit the data for ${\cal O}_{\rm max}(T,L)$ to the above form using $T_c$, $\nu$,
$a_0	$, $a_1$, $a_2\dots$ as free fitting parameters.  Varying the system sizes $L$ and
temperature window $|T-T_c|$ of the data used in the fit, as well as varying the order
of the above polynomial expansion, will give confidence on the significance 
of the fit.  

Having obtained the value of $T_c$, plotting $L_{z \,{\rm max}}(T_c)$ vs. $L$ determines the anisotropy exponent $\zeta$ by Eq.(\ref{eLz}),
\begin{equation}
L_{z \,{\rm max}}(T_c)\sim L^\zeta\enspace.
\label{eLz2}
\end{equation}
Knowing $T_c$, $\nu$ and $\zeta$, plotting 
${\cal O}_{\rm max}(T,L)$ vs. $tL^{1/\nu}$ and 
${\cal O}(T_c,L,L_z)$ vs. $L_z/L^\zeta$ should collapse 
the respective data to a single scaling curve.

\subsection{Observables}

To carry out the scaling analysis outlined in the previous section, we now have to determine appropriate observables to measure.  

For the 3D XY model of Eq.(\ref{eHXY}), we expect below $T_c$ a non-vanishing order parameter, $\psi=(1/N)\sum_i{\rm e}^{i\theta_i}$.  We define the real part of $\psi$ as,
\begin{equation}
M={1\over N}\sum_i\cos\theta_i\enspace,
\label{eM}
\end{equation}
and construct its Binder ratio\cite{Binder}
\begin{equation}
g(T,L,L_z)\equiv 1-\left[{\langle M^4\rangle\over 3\langle M^2\rangle^2}\right]=\tilde g(tL^{1/\nu}, L_z/L^\zeta)\enspace.
\label{eg}
\end{equation}
Because the scaling dimension of $M$ cancels in taking the ratio above,
the Binder ratio $g$ has scaling dimension zero, and so has the scaling form of Eq.(\ref{eScale2}).
In the above, $\langle\dots\rangle$ denotes the usual thermal average, while $[\dots]$ denotes the average over different realizations of the columnar disorder.  In the denominator of Eq.(\ref{eg}), the square of the expectation value is evaluated using two replicas with identical disorder, indexed by $a$ and $b$,
$\langle M^2\rangle^2\equiv\langle (M^a)^2\rangle\langle (M^b)^2\rangle$, in order to avoid bias.\cite{Olson}

Another observable we have measured is obtained by considering the dependence of the total free energy on the total applied twist across the system.\cite{Reger} 
Sensitivity to boundary conditions, in this case specified by the twist $\Delta_\mu$ in Eq.(\ref{eFTBC}),  is one of the signatures of an ordered phase.  The XY model is therefore 
phase coherent when the total free energy ${\cal F}$ varies with twist $\Delta_\mu$.
${\cal F}$ is computed from a partition function sum over the $\theta_i^\prime$ using
the Hamiltonian ${\cal H}_{\rm XY}[\Delta_\mu]$ of Eq.(\ref{eHXY2}).  A convenient measure of the dependence of
${\cal F}$ on $\Delta_\mu$ is obtained by looking at the curvature of ${\cal F}(\Delta_\mu)$
at its minimum.  In Appendix A we show that this minimum always  occurs at $\Delta_\mu=0$.
We therefore consider,
\begin{equation}
\left.{\partial^2{\cal F}\over\partial\Delta_\mu^2}\right|_{\Delta_\mu=0} =
\left\langle{\partial^2{\cal H}_{\rm XY}\over\partial\Delta_\mu^2}\right\rangle-
{1\over T}\left\langle\left({\partial{\cal H}_{\rm XY}\over\partial\Delta_\mu}\right)^2\right\rangle\enspace,
\label{edFdD}
\end{equation}
where ${\cal H}_{\rm XY}$ is that of Eq.(\ref{eHXY2}), and the averages on the right
hand side are taken in the ensemble with $\Delta_\mu=0$.  

Since the total free energy ${\cal F}$ and the total twist $\Delta_\mu$ are both scale
invariant quantities, then $\partial^2{\cal F}/\partial\Delta_\mu^2$ has scaling dimension zero.
These derivatives are usually defined in terms of the {\it helicity modulus}\cite{Teitel}
$\Upsilon_\mu$, which is the derivative of the free energy {\it density} with respect to the twist {\it per length}.  We have
in three dimensions,
\begin{eqnarray}
\left.{\partial^2{\cal F}\over\partial\Delta_z^2}\right|_{\Delta_\mu=0}&=&{L^2\over L_z}\Upsilon_z\enspace, \\
\left.{\partial^2{\cal F}\over\partial\Delta_x^2}\right|_{\Delta_\mu=0}&=&L_z\Upsilon_x\enspace.
\label{eUps}
\end{eqnarray}
Averaging over the disorder, 
we find that, for fixed $T$ and $L$, $(L^2/L_z)[\Upsilon_z]$ decreases monotonically as $L_z$
increases, while $L_z[\Upsilon_x]$ increases monotonically as $L_z$ increases.  In order to have 
an observable which has a maximum as a function of $L_z$, we therefore consider the
product,
\begin{equation}
L^2[\Upsilon_x\Upsilon_z]=\tilde u(tL^{1/\nu},L_z/L^\zeta)\enspace,
\label{eUps2}
\end{equation}
which has the same scaling form as Eq.(\ref{eScale2}).

\subsection{Monte Carlo Methods and Error Estimation}

\begin{figure}[!]
\resizebox{!}{8cm}{\includegraphics{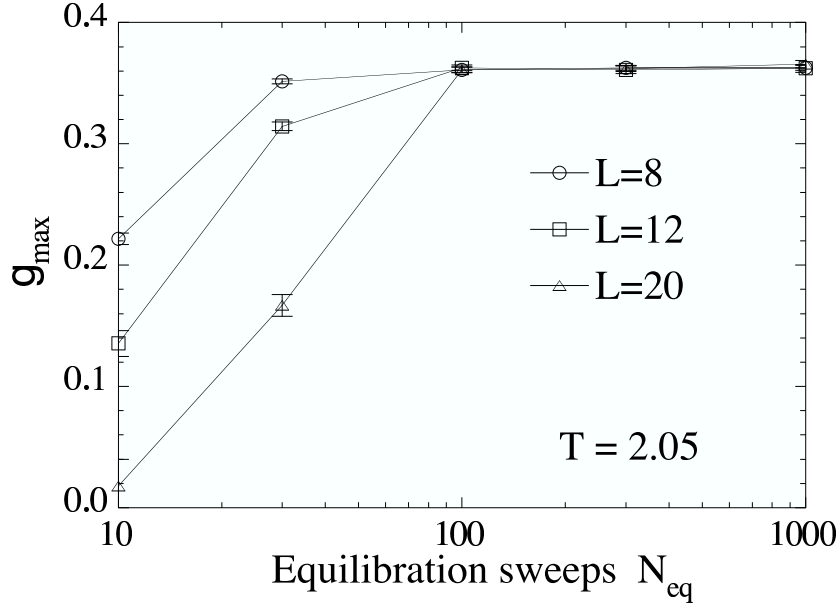}}
\caption{Binder ratio maximum for the 3D XY model, $g_{\rm max}$ at $T=2.05\simeq T_c$,  computed with $N_{\rm eq}$ equilibration sweeps followed by $N_{\rm eq}$ sweeps to compute averages.
$N_{\rm eq}=100$ is sufficient for good equilibration for all sizes $L$.}
\label{f1}
\end{figure}

In order to achieve accurate results, averaging over many disorder realizations for many different aspect ratios, $L_z/L$, it is essential to have an efficient simulation algorithm.  For equilibrium simulation of the 3D XY model, the lack of frustration allows us to use the Wolff\cite{Wolff} cluster algorithm to avoid critical slowing down.  We typically use 100 Wolff sweeps to approach equilibrium, followed by 200 Wolff sweeps to compute averages; one Wolff sweep is defined as building clusters until each phase angle $\theta_i$ has been updated once on average.  Between $3000$ and $5000$ different realizations of the random disorder are averaged over near the critical point, with fewer realizations used away from the critical point.
A test of the equilibration of our simulations is shown in Fig.~\ref{f1}, where we see that the above simulation lengths are sufficient.  

To estimate the statistical error in our results we use the following method.  For
our raw data, our average is just the average over the individual values obtained
in $N_d$ independent realizations of the random disorder.  Our estimated error
is determined from the standard deviation $\sigma$ of these $N_d$ independent values,
${\rm error}=\sigma/\sqrt{N_d}$.
%
%
To estimate the statistical error in the fitting parameters of our finite size scaling 
analysis, we take the following approach.  From our original data set we construct many 
(typically $1000$) {\it fictitious} data sets by adding to each data point a random Gaussian variable with zero mean, and standard deviation equal to the estimated statistical error of the data point.
We then fit each of the fictitious data sets.  The standard deviation of the values of the resulting
fitting parameters then gives our estimate of the statistical error in the fitting parameter.

Harder to estimate are the possible systematic errors in our results.  Here we rely on
varying parameters of our analysis, such as the order of a polynomial fit, or the
system sizes $L$ used in the fit, in order to get a feeling for the likely accuracy of
our results.

\subsection{Results}

\begin{figure}[!]
\resizebox{!}{8cm}{\includegraphics{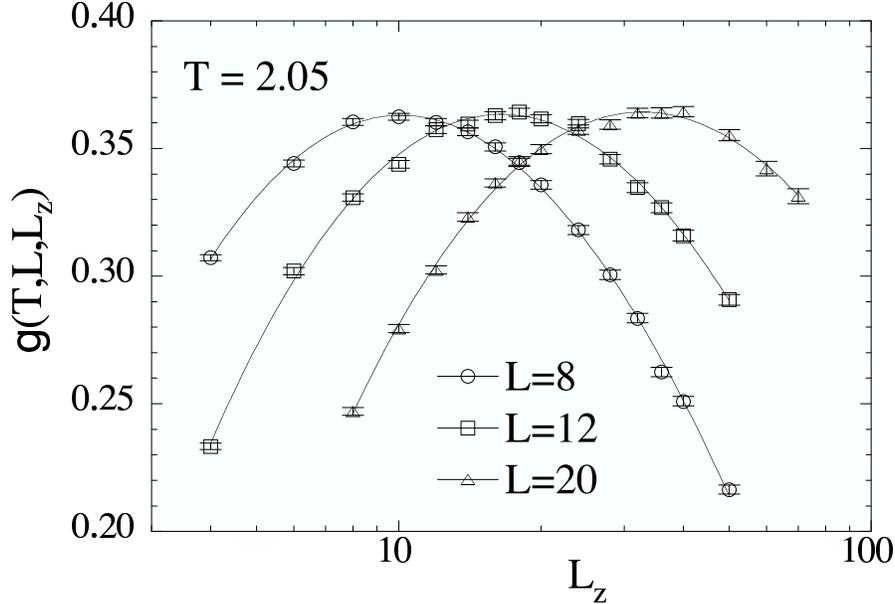}}
\caption{Binder ratio, $g(T,L,L_z)$ vs. $L_z$ for several values $L$ at the fixed $T=2.05$.  Solid curves are cubic polynomial fits in $\ln L_z$.  }
\label{f3}
\end{figure}

We now present our results from simulations of the XY model of Eq.(\ref{eHXY}).
In Fig.~\ref{f3} we plot our results for the Binder ratio of Eq.(\ref{eg}), 
$g(T,L,L_z)$ vs. $L_z$, for sizes $L=8,12,20$, at the fixed temperature $T=2.05$.  We see 
that for each $L$, $g(T,L,L_z)$ has a clear maximum at a particular $L_{z\,{\rm max}}$.
Note that the maximum values of these curves appear to be equal for the different values of $L$.  From Eq.(\ref{eOmax}) we therefore infer that the temperature $T=2.05$ is 
approximately the critical temperature $T_c$.
To determine the precise values of $L_{z\,{\rm max}}$ and the maximum values 
$g_{\rm max}(T,L)=g(T,L,L_{z\,{\rm max}})$, we fit the data for each $L$ to a cubic
polynomial in $\ln L_z$ (these are the solid curves in Fig.~\ref{f3}).
The $L_{z\,{\rm max}}$ obtained this way are not, in general, integer values.
We have also done such fits using a quadratic polynomial in $\ln L_z$; the
difference in values obtained from the cubic vs. the quadratic fit provides our
estimate of the systematic error of this procedure.
We find that for $g_{\rm max}(T,L)$ this systematic 
error is always smaller than the estimated statistical error of the cubic fits;
for $L_{z\,{\rm max}}$ the systematic error is bigger.  
This reflects the simple fact that $g(T,L,L_z)$, being a maximum with
zero slope, varies only quadratically with deviations from the true $L_{z\,{\rm max}}$,
and hence may be determined more accurately.  Henceforth, the error bars we
use for $g_{\rm max}$ are the above estimated statistical errors, while
the error bars we use for $L_{z\,{\rm max}}$ are the above defined systematic
errors.
\begin{figure}[!]
\resizebox{!}{8cm}{\includegraphics{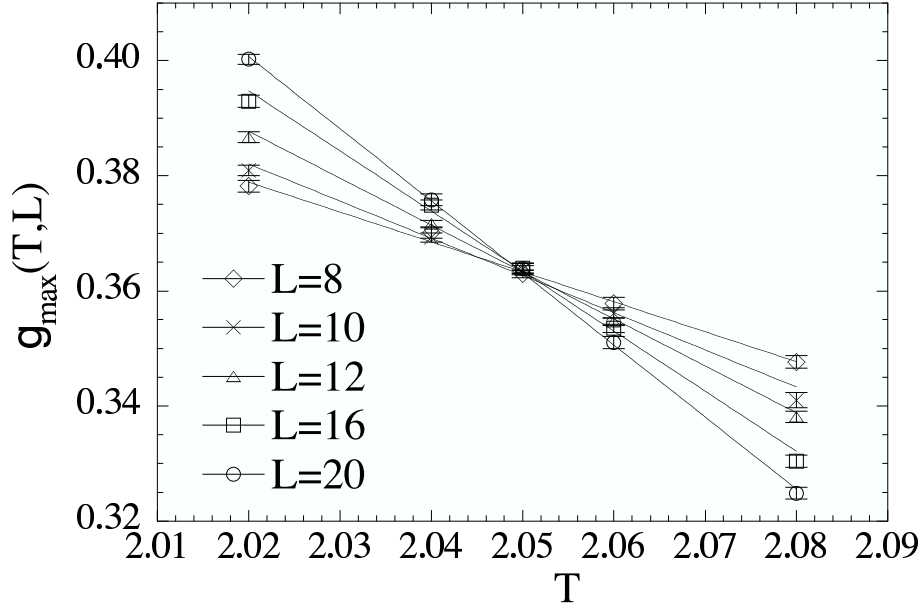}}
\caption{Binder ratio maximum $g_{\rm max}(T,L)$ vs. $T$ for various system sizes $L$.  
The common intersection determines $T_c\simeq 2.05$. Solid lines are linear fits to the data.}
\label{f4}
\end{figure}

\begin{figure}[!]
\resizebox{!}{8cm}{\includegraphics{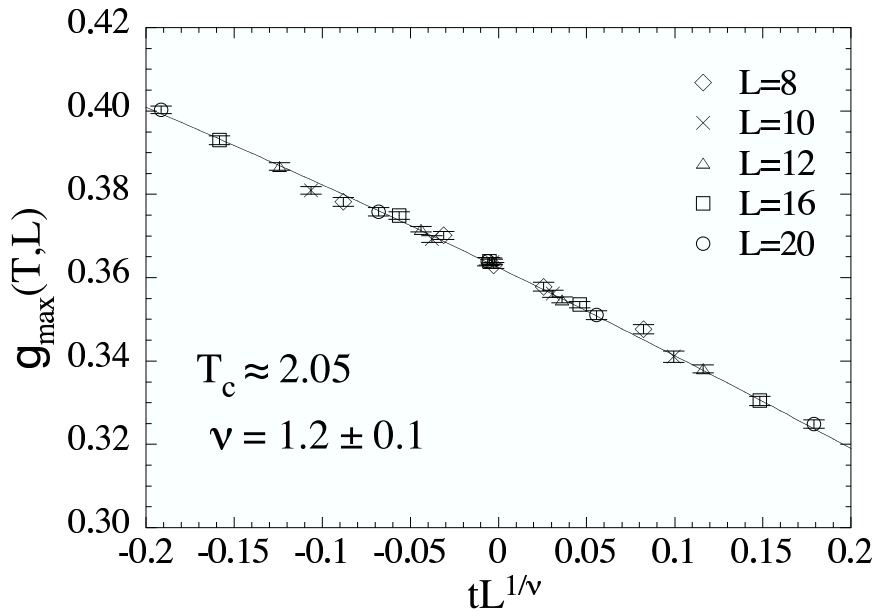}}
\caption{Scaling collapse of $g_{\rm max}(T,L)$ vs. $((T-T_c)/T_c)L^{1/\nu}$.
The values of $T_c=2.051$ and $\nu=1.18$ from the last row of Table~\protect\ref{t1}
are used. The solid curve is the fitted cubic polynomial.}
\label{f5}
\end{figure}

Proceeding in this way at other temperatures, we plot in Fig.~\ref{f4} the values of $g_{\rm max}(T,L)$ vs. $T$ for $L=8$, $10$, $12$, $16$ and $20$.  The different curves all intersect at a common point, determining $T_c\simeq 2.05$.
To determine the correlation length exponent $\nu$, and get a more precise
estimate of $T_c$, we now fit the data of Fig.~\ref{f4} to a polynomial
expansion as in Eq.(\ref{epoly}).  In Table~\ref{t1} we show the results 
from both quadratic and cubic polynomial fits, using different system sizes
$L$; we systematically drop the smallest sizes since scaling holds only in
the asymptotic large $L$ limit.  Our results give a consistent value of $T_c\simeq 2.05$.
The values of $\nu$ that we obtain are consistent within the estimated statistical errors,
however we see a small systematic increase in the value of $\nu$ when we restrict the
data to larger system sizes.  We therefore estimate $\nu = 1.2\pm 0.1$.
Note that our value $\nu>1$, satisfies 
the Chayes lower bound condition,\protect\cite{Chayes}
as generalized\cite{Wallin3} for correlated disorder, $\nu > 2/d^*$, where
$d^*=2$ is the number of dimensions in which the system is disordered.
In Fig.~\ref{f5} we replot the data of Fig.~\ref{f4} in a scaled form, $g_{\rm max}(T,L)$
vs. $((T-T_c)/T_c)L^{1/\nu}$.  We use the value of $T_c$ and $\nu$ from Table~\ref{t1}
for the cubic fit to sizes $L=12-20$.  The solid line is the fitted cubic polynomial.
As is seen, the data collapse is excellent.

\begin{figure}[!]
\resizebox{!}{8cm}{\includegraphics{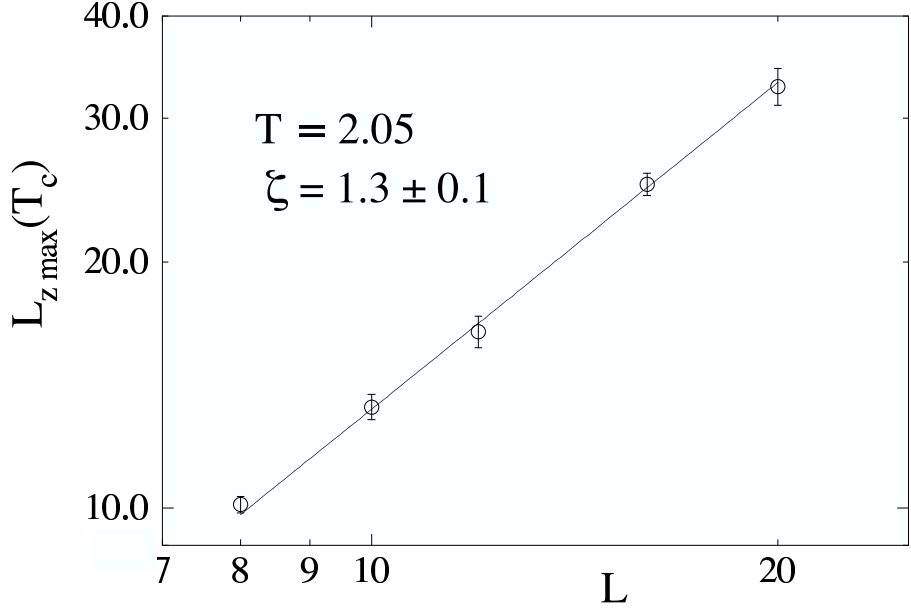}}
\caption{Log-log plot of $L_{z\,{\rm max}}$ vs. $L$, at $T=2.05\simeq T_c$. 
The solid straight line is the best power law fit using the data for $L=10 - 20$
and yields the value $\zeta=1.329\pm0.08$ (see Table~\protect\ref{t2}).}
\label{f6}
\end{figure}

\begin{figure}[!]
\resizebox{!}{8cm}{\includegraphics{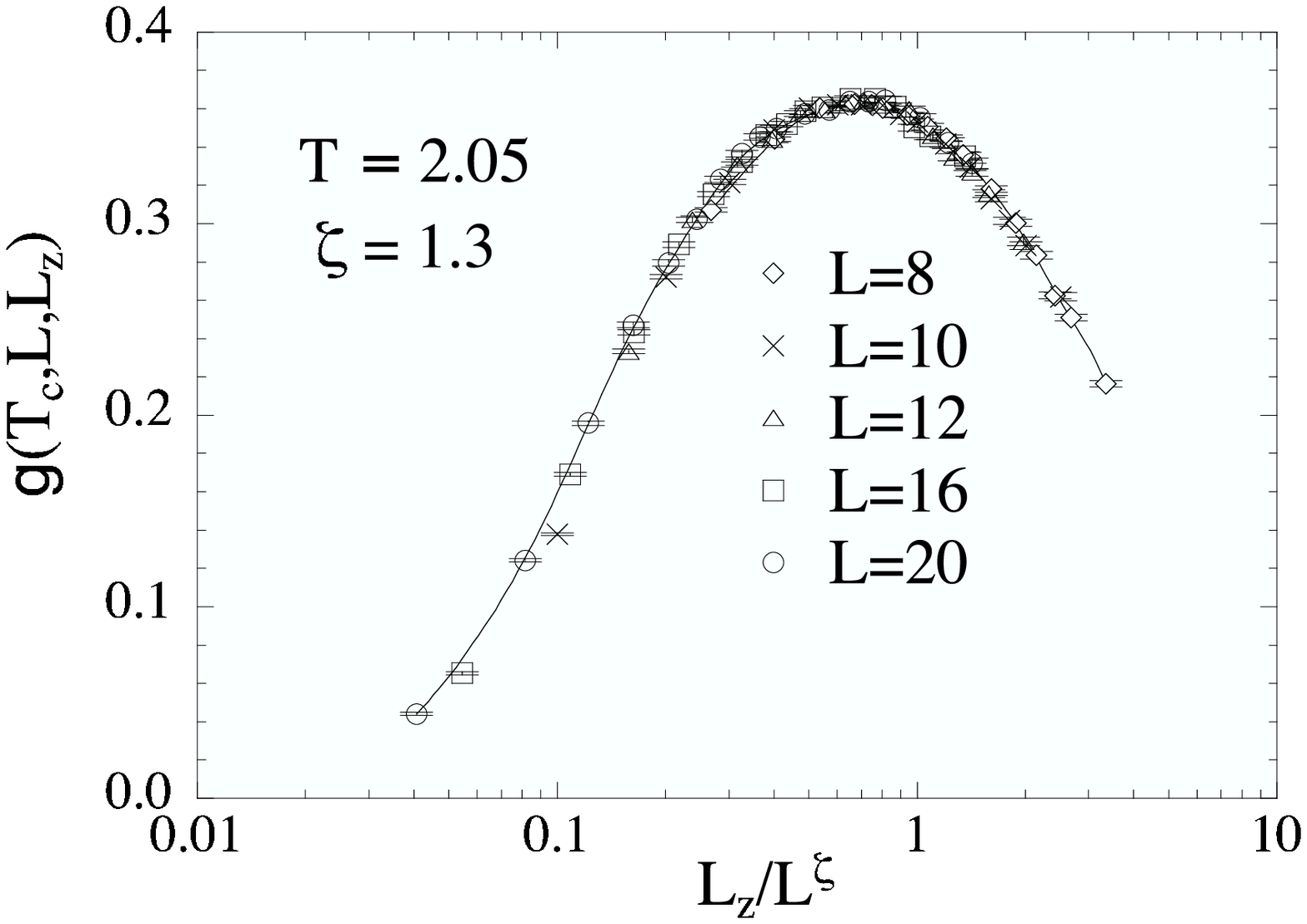}}
\caption{Scaling collapse of $g(T_c,L,L_z)$ vs. $L_z/L^{\zeta}$, for data
at $T=2.05\simeq T_c$, using $\zeta=1.3$. The solid line is a guide to the eye only.}
\label{f8}
\end{figure}

\begin{table}
  \centering 
  \begin{ruledtabular}
  \begin{tabular}{c|c|c|c}
$L$ & order & $T_c$ &   $\nu$   \\\hline
$8 - 20$ & quadratic & $2.052 \pm 0.001$ & $1.07 \pm 0.03 $ \\
             & cubic & $2.052 \pm 0.001$ & $1.04 \pm 0.06 $ \\\hline
$10 - 20$ & quadratic & $2.052 \pm 0.001$   & $1.10 \pm  0.04 $  \\ 
               & cubic & $2.052 \pm 0.001$  & $1.10 \pm 0.06$ \\\hline
$12 - 20$ & quadratic & $2.051 \pm 0.001$   & $1.18 \pm  0.04 $   \\ 
             & cubic & $2.051 \pm 0.001$  & $1.18 \pm 0.05 $ \\
\end{tabular}
\end{ruledtabular}
  \caption{Fitting parameters $T_c$ and $\nu$ from quadratic and cubic polynomial
  scaling fits to the data of Fig.~\protect\ref{f4}. Results for different ranges of 
 system sizes $L$ are shown.}
  \label{t1}
\end{table}
Having found the value of $T_c$, we next determine the anisotropy exponent $\zeta$.
In Fig.~\ref{f6} we show a log-log plot of our data for $L_{z\,{\rm max}}$ vs. $L$,
at the temperature $T=2.05\simeq T_c$.   Fitting to Eq.(\ref{eLz2}), $L_{z\,{\rm max}}\sim L^\zeta$, we get the results summarized in Table~\ref{t2} for different ranges of system 
sizes $L$.  The results are consistent within the estimated statistical error and we find
$\zeta = 1.3\pm 0.1$.
To check the consistency of our value for $\zeta$,  in
Fig.~\ref{f8} we plot 
$g(T_c,L,L_z)$ vs. $L_z/L^{\zeta}$, using our data at $T=2.05$ and the
above determined value of $\zeta=1.3$.
As expected from Eq.(\ref{eScale2}), the data for the different values of
$L$ and $L_z$ show a very good collapse to a single scaling curve.

\begin{table}
  \centering 
  \begin{ruledtabular}
  \begin{tabular}{c|c|c|c}
$L_{\rm min}$ & $8$ & $10$ &   $12$   \\\hline
$\zeta$ & $1.29\pm 0.05$ & $1.33\pm 0.08$ & $1.37\pm 0.12$ \\
\end{tabular}
\end{ruledtabular}
  \caption{Anisotropy exponent $\zeta$ from power law fits, $L_{z\,{\rm max}}\sim L^\zeta$
   to system sizes $L=L_{\rm min} - 20$.}
  \label{t2}
\end{table}
%

\begin{figure}[htb]
\resizebox{!}{8cm}{\includegraphics{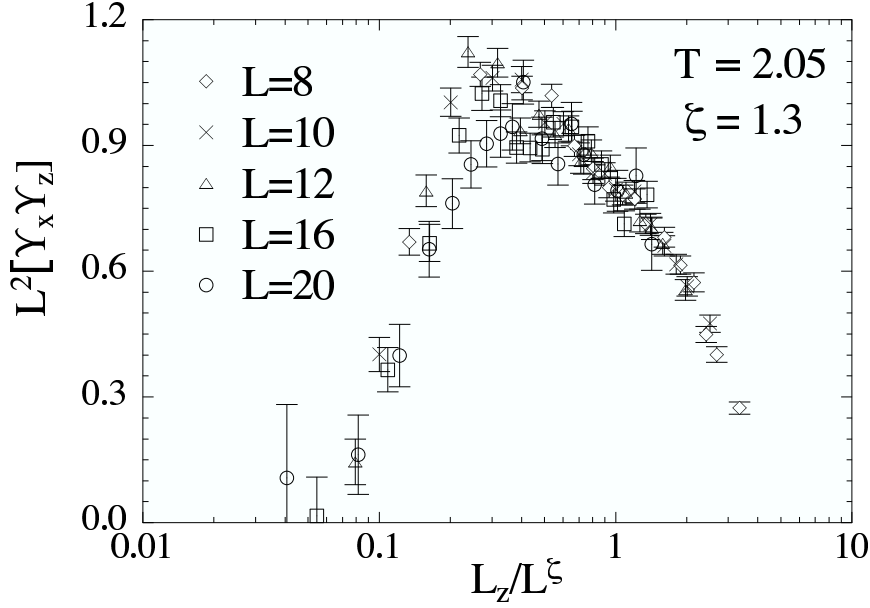}}
\caption{Attempted scaling collapse of $L^2[\Upsilon_x\Upsilon_z]$ vs. $L_z/L^\zeta$.
Data is for $T=2.05\simeq T_c$, using $\zeta=1.3$.}
\label{f10.5}
\end{figure}

We have also tried a similar scaling analysis for the helicity moduli product, $L^2[\Upsilon_x\Upsilon_z]$, of Eq.(\ref{eUps2}).  However here we have found
less satisfactory results. We find that for a given system size $L$, the $L_{z\,{\rm max}}$
where $L^2[\Upsilon_x\Upsilon_z]$ has its maximum occurs at a smaller value of
$L_z$ than was the case for the Binder ratio $g_{\rm max}$.  Such smaller system
sizes presumably have larger corrections to scaling.  We have also found the statistical
error in $L^2[\Upsilon_x\Upsilon_z]$ to be larger than we found for $g_{\rm max}$,
possibly because the Binder ratio $g$ involves a ratio between fluctuating
quantities and so has smaller sample to sample fluctuations.\cite{Huse}
As a consequence of these two effects, we could not arrive at a convincing
determination of $T_c$ and $\nu$ from the $L^2[\Upsilon_x\Upsilon_z]$ data
alone.  However, to illustrate our results we can make use of the values of
$T_c\simeq 2.05$ and $\zeta\simeq 1.3$ found in our analysis of $g_{\rm max}$.
In Fig.~\ref{f10.5} we therefore show a scaling collapse similar to that of Fig.~\ref{f8},
plotting $L^2[\Upsilon_x\Upsilon_z]$ vs. $L_z/L^\zeta$, using our
data at $T=2.05\simeq T_c$ and the above value of $\zeta$.

We see clearly in Fig.~\ref{f10.5} the above effects: error bars are considerably
larger than in Fig.~\ref{f8}, and the peak is at a smaller value of $L_z/L^\zeta$.
The scaling collapse is not bad for the bigger systems sizes, corresponding to 
larger values of $L_z/L^\zeta$.  However it is rather scattered near the peak and below it.
We conclude that it would be necessary to average over many more disorder
realizations to reduce the errors, and perhaps also go to larger system sizes, 
in order to get a convincing scaling analysis from the helicity product
$L^2[\Upsilon_x\Upsilon_z]$ on its own.

\section{Dynamic Critical Behavior}

As one approaches the critical temperature $T_c$, we expect relaxation times to diverge as
$\tau\sim\xi^z$, defining the dynamic critical exponent $z$.  To compute equilibrium critical exponents, it is sufficient that the simulation dynamics satisfies detailed balance; the details of the dynamics are otherwise irrelevant.  
Thus the exact duality between ${\cal H}_{\rm XY}$ and ${\cal H}_{\rm loop}$
implies that the unscreened and the strongly screened superconductors have the
same {\it equilibrium} critical behavior.
For the dynamic critical behavior, however, the value of 
$z$ will in general depend on the details of the dynamics,\cite{Hohenberg}
and some works suggest that it may even vary for different types of relaxational dynamics or
different boundary conditions.\cite{Minnhagen, Minnhagen2}  
There is thus no reason, {\it a priori}, to expect the same dynamic critical behavior for
the XY model, expressed in terms of a dynamical rule for the phase variables
$\theta_i$, as compared to the loop model, expressed in terms of a dynamical
rule for the vortex line variables $n_{i\mu}$.  In this section, therefore, we will 
present results from explicit simulations of the loop model as well as the XY model.

Because the true dynamics of a superconductor is local, it is not physically meaningful to compute the dynamic critical exponent within accelerated global algorithms such as the Wolff algorithm, which we used to compute equilibrium properties.  We therefore will use a {\it local} Monte Carlo dynamics for both ${\cal H}_{\rm XY}$ and ${\cal H}_{\rm loop}$.  
Even within such local algorithms, it is not obvious how universal the dynamical
critical behaviors may be.  Thus it is unclear that our results will correspond to
what is seen in experiments.  Nevertheless it will be interesting to see if the
two models give similar or different values of $z$.

The relative loss of efficiency that results  from using such local algorithms 
means that we will be unable to do
as extensive an exploration of the parameter space as we did for our equilibrium
analysis.  But this is not necessary.  We can make use of our already obtained equilibrium results, and simulate only at the value of $T=T_c$, using system aspect ratios
$L_z=\gamma L^\zeta$.  For our simulation of ${\cal H}_{\rm loop}$, we
will simulate the loop model which is exactly dual (see Appendix A, Eq.(\ref{eIn}))
to the cosine XY model that we have used in
our equilibrium simulations, so as to make use of these known values of $T_c$ 
and $\zeta$.

\subsection{Monte Carlo Methods and Scaling}

For the XY model of an unscreened superconductor we use a standard single spin heat bath algorithm, with fixed periodic boundary conditions on the $\theta_i$.  In this algorithm, 
a phase angle
$\theta_i$ is selected at random and replaced with a new randomly chosen $\theta_i^\prime$.
This update attempt is then accepted with probability $1/[1+{\rm exp}(\Delta E/T)]$
where $\Delta E$ is the change in energy.
One sweep, consisting of $N=L^2L_z$ update attempts, is taken as one time step, $\Delta t=1$.  We average over $300-700$ disorder realizations depending on system size $L$.

For the loop model of a strongly screened superconductor, we again use a heat bath algorithm
in which the attempted excitation consists of an elementary vortex loop circulating about a
randomly chosen plaquette of the grid.  Adding only such closed loop excitations
corresponds to the ensemble in which the average internal magnetic field is constrained to ${\bf B}=0$ (see Appendix A).   One sweep, consisting of $3N$ such update
attempts, is taken as one time step, $\Delta t=1$.  We average over
$1000 - 2000$ disorder realizations depending on system size $L$.

In general, we expect the relaxation time $\tau$ to obey the scaling equation,
\begin{equation}
\tau(T,L,L_z)=b^z\tilde\tau (tb^{1/\nu},L/b,L_z/b^\zeta)\enspace,
\label{etau}
\end{equation}
where $b$ is an arbitrary length rescaling factor.  For $b=L$, $T=T_c$, and
$L_z=\gamma L^\zeta$, this reduces to the simple,
\begin{equation}
\tau \sim L^z \enspace.
\label{etau3}
\end{equation}
For both the XY model and the loop model, we simulate with values of
$L_z=\gamma L^\zeta$ as determined by the fit shown in Fig.~\ref{f6}.
For the XY model, to approximate
non-integer values of $L_z$ we use linear interpolation of simulation data for the two closest integer values of $L_z$.  For the loop model we simply
use results from the closest integer value of $L_z$.

\subsection{Observables}

\subsubsection{XY Model}

For the XY model we have tried two independent methods of determining $z$, analogous
to the two quantities $g$ and $L^2[\Upsilon_x\Upsilon_z]$ used in our equilibrium
simulations.  The first is to look at the decay of correlations in the order parameter $M$
of Eq.(\ref{eM}), defining the relaxation time $\tau$ by,
\begin{equation}
\tau=1+2\sum_{t=1}^{t_0}\left[{\langle M(t)M(0)\rangle\over\langle M^2\rangle}\right]
\sim L^z\enspace,
\label{etau4}
\end{equation}
where $t_0$ is chosen large enough so that $\tau$ is independent of $t_0$.  The
ratio in the above ensures that the quantity being summed over has scaling dimension zero, 
and hence the sum scales as $\tau\sim L^z$.

The second method is to look at correlations of the supercurrent $I_\mu$, defined by,
\begin{equation}
I_\mu=\left.{\partial{\cal H}_{\rm XY}\over\partial\Delta_\mu}\right|_{\Delta_\mu=0}
={1\over L_\mu}\sum_i J_{i\mu}\sin(\theta_{i+\hat\mu}-\theta_i)\enspace.
\label{eI}
\end{equation}
In terms of $I_\mu$ one can define the conductance in the $\hat\mu$ direction
by the Kubo formula,\cite{Minnhagen}
\begin{equation}
G_\mu={1\over 2T}\sum_{t=-t_0}^{t_0}\Delta t\left[\langle I_\mu (t)I_\mu (0)\rangle\right]
\sim L^z\enspace,
\label{eGmu}
\end{equation}
where again $t_0$ is chosen large enough that $G_\mu$ is independent of $t_0$.
Since $I_\mu=\left. (\partial{\cal H}_{\rm XY}/\partial\Delta_\mu)\right|_{\Delta_\mu=0}$,
and ${\cal H}_{\rm XY}$ and $\Delta_\mu$ are scale invariant, then $I_\mu$, and hence
the correlation summed over in in the definition of $G_\mu$, has scaling dimension zero.
Therefore, the sum which defines $G_\mu$ scales as $\tau\sim L^z$.

\subsubsection{Loop Model}

For the loop model we consider the total resistance, defined as follows.\cite{Wallin1,Wallin3}  
Let $Q_\mu(t)$ be the total projected loop area with normal in direction
$\hat\mu$ at simulation time $t$.
Each time an oriented elementary vortex loop with normal in direction $\pm\hat\mu$
is accepted, $Q_\mu$ changes by $\pm 1$.  Let $\Delta Q_\mu(t)\equiv Q_\mu(t)-Q_\mu(t-1)$ be
the total change in this area after one sweep through the entire system; each sweep represents
$\Delta t=1$.  In one such sweep, the total average phase angle change across the
length of the system (in the dual screened XY superconductor model) in direction $\hat\mu$ is just $2\pi\Delta Q_\mu/L_\nu L_\sigma$, 
where $\mu$, $\nu$, $\sigma$ are a cyclic permutation of $x$, $y$, $z$.
By the Josephson relation, the total voltage drop
across the system in direction $\hat\mu$ will then be
\begin{equation} 
V_\mu(t)=\left({\hbar\over 2e}\right) \left({2\pi\over L_\nu L_\sigma}\right) \left({\Delta Q_\mu \over \Delta t}\right)=
\left({h\over 2e}\right) \left({1\over L_\nu L_\sigma}\right) \left({\Delta Q_\mu\over \Delta t}\right)\enspace.  
\label{eVt}
\end{equation}
Henceforth we define our units of voltage such that $h/2e\equiv 1$.
We then define the total resistance  in direction $\hat\mu$  by the Kubo formula,\cite{Young2}
\begin{equation}
R_\mu={1\over 2T}\sum_{t=-t_0}^{t_0}\Delta t[\langle V_\mu(t)V_\mu(0)\rangle]\enspace,
\label{eR}
\end{equation}
where again $t_0$ is chosen large enough so that $R_\mu$ is independent of $t_0$.
Since the total voltage drop $V_\mu$ is the time rate of change of the total phase angle difference 
across the system, and since the total phase angle difference is a scale invariant quantity,
we have the scaling $V_\mu\sim 1/\tau$.  Thus the resistance above scales as,
\begin{equation}
R_\mu\sim 1/\tau\sim L^{-z}\enspace.
\label{eR2}
\end{equation}

\subsection{Results}

\subsubsection{XY Model}

\begin{figure}[htb]
\resizebox{!}{8cm}{\includegraphics{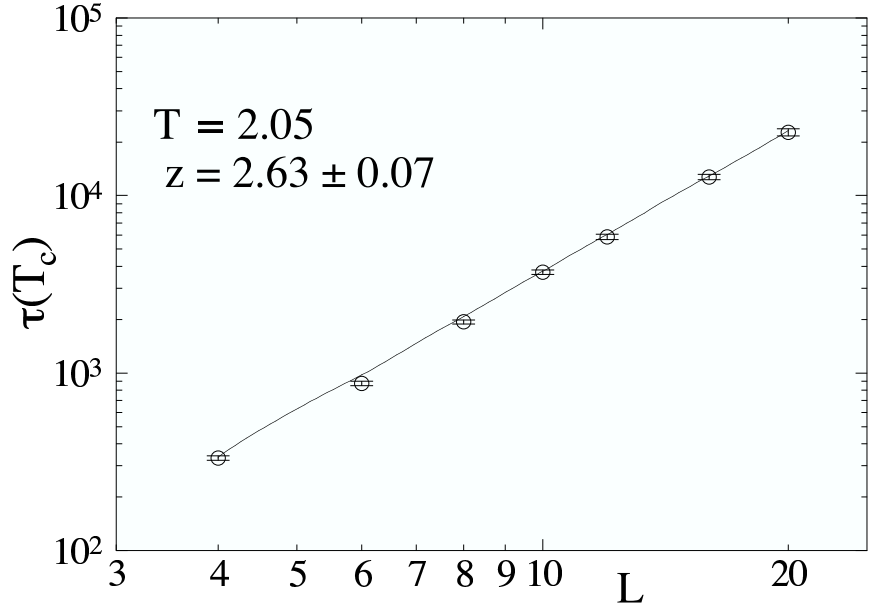}}
\caption{Log-log plot of order parameter relaxation time $\tau$ of Eq.(\ref{etau4}) vs. system size $L$,
for $T=2.05\simeq T_c$ and $L_z\sim L^\zeta$.  Solid line is the best power law
fit for sizes $L=10 - 20$, and determines $z=2.63\pm 0.07$ (see Table~\protect\ref{t3}).}
\label{fx1}
\end{figure}

In Fig.~\ref{fx1} we show a log-log plot of our results for the order parameter relaxation 
time $\tau$ of Eq.(\ref{etau4}) vs. system size $L$, for $T=2.05\simeq T_c$
and $L_z\sim L^\zeta$.  Our results are obtained using $5\times 10^5$ MC sweeps to 
equilibrate, followed by $10^6$ sweeps to compute averages.  
Fitting to the power law, $\tau\sim L^z$, we get the results summarized
in Table~\ref{t3}, for different ranges of systems size $L$.  The results are consistent
within the estimated statistical error, with a small systematic tendency to lower values 
as we restrict the fitted data to larger system sizes.  We find $z_{\rm XY}=2.63\pm 0.07$.

\begin{table}
  \centering 
  \begin{ruledtabular}
  \begin{tabular}{c|c|c|c|c}
$L_{\rm min}$ & $6$ & $8$ & $10$ &   $12$   \\\hline
$z_{\rm XY}$ & $2.72\pm 0.04$ &$2.69\pm 0.05$ & $2.63\pm 0.07$ & $2.60\pm 0.03$ \\
\end{tabular}
\end{ruledtabular}
  \caption{Dynamic exponent $z_{\rm XY}$ from power law fits, $\tau\sim L^z$,
   to system sizes $L=L_{\rm min} - 20$.}
  \label{t3}
\end{table}

\begin{figure}[htb]
\resizebox{!}{8cm}{\includegraphics{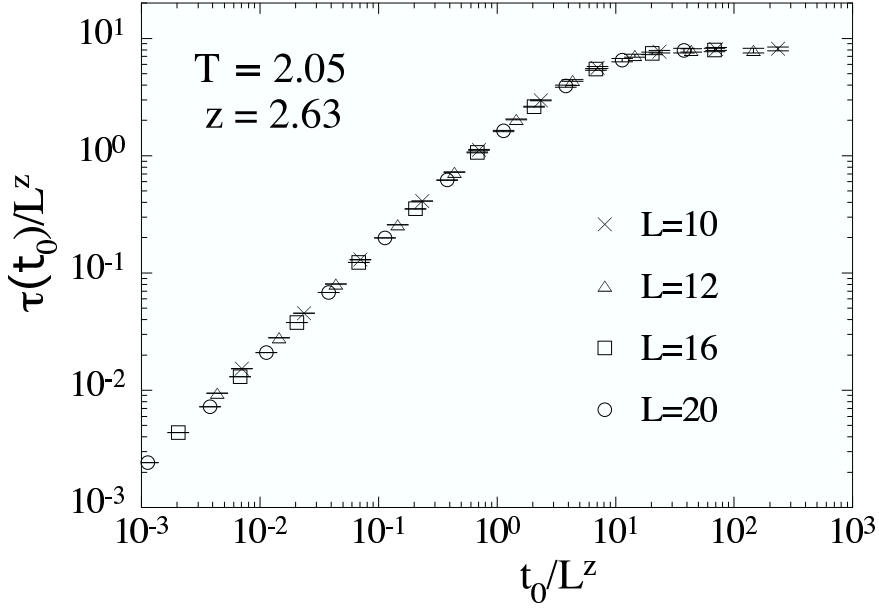}}
\caption{Log-log scaling plot of order parameter relaxation time $\tau(t_0)/L^z$ vs. $t_0/L^z$ for $T=2.05\simeq T_c$,
$L_z\sim L^\zeta$, and various values of $L$.  Using $z_{\rm XY}=2.63$ gives an
excellent collapse for the entire range of $t_0$.}
\label{fx2}
\end{figure}

As another check on our above determination of $z_{\rm XY}$, we
consider the following.  In principal, $\tau$ is defined by taking
$t_0$ in Eq.(\ref{etau4}) sufficiently large so that $\tau$ is independent of $t_0$; our data in Fig.~\ref{fx1}
satisfies this condition.  How big $t_0$ must be for this to happen is set by the time
scale $\tau$ itself.  Therefore, we expect that if we compute $\tau$ for arbitrary $t_0$,
then $\tau(t_0)$ should scale as,
\begin{equation}
  \tau (t_0)\sim L^z\tilde\tau (t_0/\tau)\sim L^z\tilde\tau(t_0/L^z)\enspace.  
  \label{ett0}
\end{equation}
In Fig.~\ref{fx2} we show a log-log plot of
$\tau(t_0) /L^z$ vs. $t_0/L^z$ for various sizes $L$ (again using $T=2.05\simeq T_c$
and $L_z\sim L^\zeta$).  Choosing the value $z_{\rm XY}=2.63$ obtained from the fit
in Fig.~\ref{fx1} we find an excellent collapse of all the data.
For large $t_0/L^z$ we see that the curve does indeed saturate to a finite constant as expected, 
however the collapse holds for the entire range of $t_0$.

\begin{figure}[htb]
\resizebox{!}{8cm}{\includegraphics{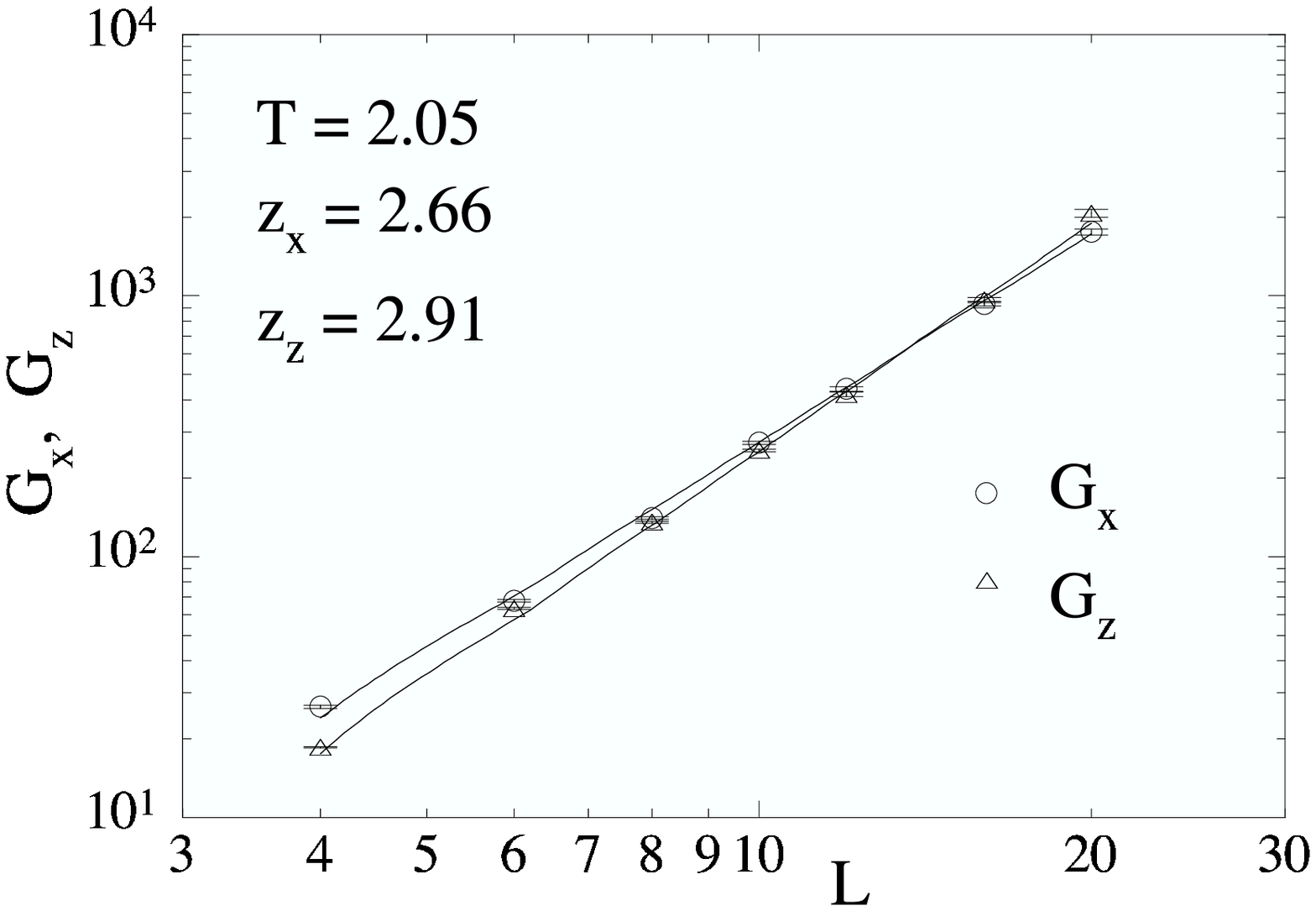}}
\caption{Log-log plot of conductances $G_x$ and $G_z$ vs. $L$, for
$T=2.05\simeq T_c$ and $L_z\sim L^\zeta$.  The fitted straight lines
determine $z_{x}=2.66$ and $z_z=2.91$.}
\label{fx3}
\end{figure}

Finally, we plot in Fig.~\ref{fx3} the conductances of Eq.(\ref{eGmu}),
$G_x$ and $G_z$ vs. $L$, for
$T=2.05\simeq T_c$ and $L_z\sim L^\zeta$.  
Our results are for $2\times 10^5$ MC sweeps to equilibrate, followed by $4\times 10^5$
sweeps to compute averages.
Fitting to the power law, $G_\mu\sim L^{z_\mu}$, we get the results summarized
in Table~\ref{t4}, for different ranges of systems size $L$.  For $G_x$ the
results $z\simeq 2.66\pm 0.04$ are consistent, within errors, with that obtained from our 
analysis of the order parameter relaxation time $\tau$.  For $G_z$, we get values for
$z$ that are somewhat larger.  However if one compares the data points for $G_x$
and $G_z$ directly, one sees that the values are all roughly equal within the 
estimated error, except for the smallest size $L=4$ (probably too small to be in
the scaling limit) and for the largest size $L=20$.  Our fit for $z$ from the
$G_z$ data is skewed by this one $L=20$ data point.  If we restrict our fit to
sizes $L=8 - 16$, we then find $z_z=2.82\pm 0.03$.  This is still somewhat 
larger than what we get from $G_x$, but within two standard deviations
of $z_x$ for the same range of sizes $L=8-16$.

\begin{table}
  \centering 
  \begin{ruledtabular}
  \begin{tabular}{c|c|c|c|c}
$L_{\rm min}$ & $6$ & $8$ & $10$ &   $12$   \\\hline
$z_{x}$ & $2.71\pm 0.02$& $2.75\pm 0.03$ &$2.66\pm 0.04$ & $2.69\pm 0.06$ \\\hline
$z_{z}$&$ 2.77\pm 0.02$ &$ 2.87\pm 0.03$ &$2.91\pm 0.04$&$3.06\pm 0.07$\\
\end{tabular}
\end{ruledtabular}
  \caption{Dynamic exponent $z_{\rm XY}$ from power law fits, $G_\mu\sim L^{z_\mu}$,
   to system sizes $L=L_{\rm min} - 20$.}
  \label{t4}
\end{table}

\subsubsection{Loop Model}

\begin{figure}[htb]
\resizebox{!}{8cm}{\includegraphics{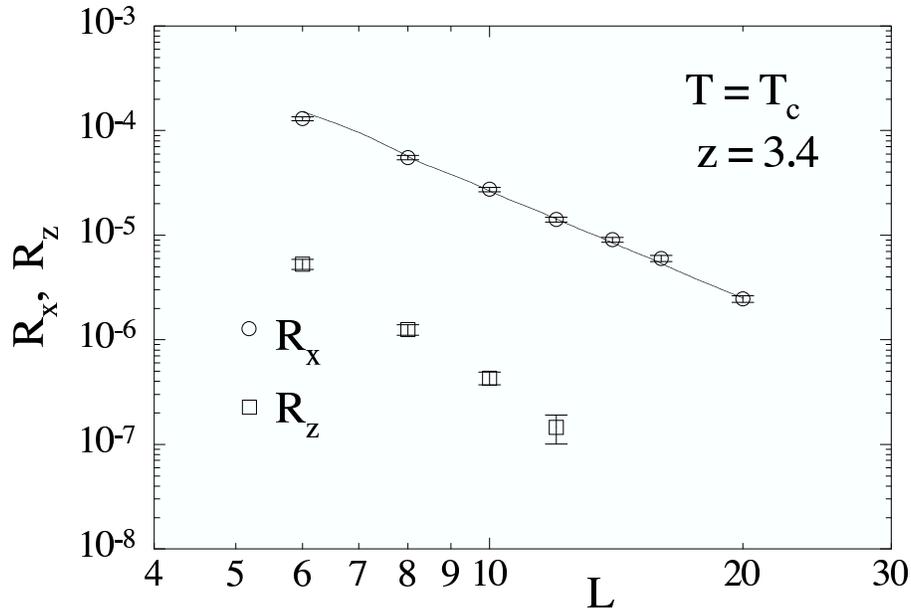}}
\caption{Log-log plot of resistance $R_x$ and $R_z$ of the loop model vs. $L$.
The solid line is the best power law fit, $R_x\sim L^{-z}$, for sizes $L=10 - 20$,
and determines the value $z_{\rm loop}=3.4\pm 0.1$ (see Table~\ref{t5}).}
\label{fx4}
\end{figure}

For our loop simulations we use the interaction of Eq.(\ref{eIn}),
exactly dual to our XY model.  This interaction is computed using
the  same distribution of $J_{i\mu}$ as we used for the XY model, 
and we simulate at the same value of $T=2.05$ as gives the critical 
point of the XY model.  We also use the same values of $L_z=\gamma L^\zeta$
as we used for the XY model, as determined from Fig.~\ref{f6}.
In Fig.~\ref{fx4} we give our results for the resistance
of the loop model, Eq.(\ref{eR}),  as a log-log plot of $R_x$
and $R_z$ vs. system size $L$.  
Our results are from $12\times 10^4$ MC sweeps to equilibrate, followed
by $24\times 10^4$ sweeps to compute averages.
Fitting to the power law of Eq.(\ref{eR2}), $R_x\sim L^{-z}$, we get the results summarized
in Table~\ref{t5}, for different ranges of systems size $L$.  The results are consistent
within the estimated statistical error, and we find $z_{\rm loop}=3.4\pm 0.1$.

\begin{figure}[htb]
\resizebox{!}{8cm}{\includegraphics{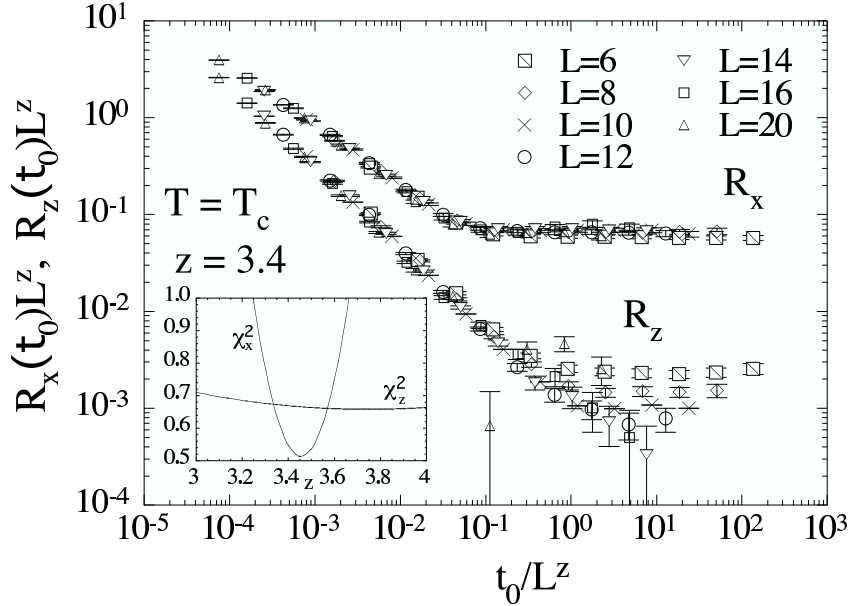}}
\caption{Log-log scaling plot of time dependent resistance $R_x(t_0)L^z$ and
$R_z(t_0)L^z$ of the loop model vs. $t_0/L^z$.  The value $z=3.4$ obtained
from the fit in Fig.~\ref{fx4} is used.  The inset gives the $\chi^2$ error of
the data collapse, as the exponent $z$ is varied. }
\label{fx5}
\end{figure}

For the case of $R_z$, parallel to the columnar defects, our simulations
were not sufficiently long to observe the necessary saturation of $R_z(t_0)$
with increasing $t_0$, except  for the smallest system sizes $L\le 12$.  We do
not believe that any estimate of $z_{\rm loop}$ based on such small system
sizes would be meaningful.  We can, however, perform the following consistency
check.  Similar to our discussion concerning $\tau(t_0)$ (see Eq.(\ref{ett0})),
we can compute $R_\mu$ of Eq.(\ref{eR}) for finite times $t_0$, and 
we expect $R_\mu(t_0)L^z$ to scale with the variable $t_0/L^z$.
In Fig.~\ref{fx5} we make such a log-log scaling plot
using the value of $z=3.4$ found for $R_x$ in Fig.~\ref{fx4}.
For $R_x$ we see that the collapse is excellent for all times $t_0$, and
the scaling curve saturates to a constant at large $t_0/L^z$ as expected.
For $R_z$, we find a good collapse for all but the largest times.  We see
that $R_z(t_0)$ saturates only for the smallest systems, and it is only here
that the collapse appears to be breaking down.  We conclude that these
small system sizes are not large enough to expect scaling for $R_z$ to hold.

We can also try to independently determine the dynamic exponent $z$ by
fitting to a data collapse as in Fig.~\ref{fx5} for all times $t_0$, rather than
just the asymptotic large time limit.  The inset to Fig.~\ref{fx5} shows
the resulting $\chi^2$ of such fits as the fitting parameter $z$ is varied.
For $R_x$, the $\chi^2$ shows a sharp minimum at $z=3.45$, in good
agreement with our earlier value of $z=3.4$ from Fig.~\ref{fx4}.
For $R_z$, the $\chi^2$ has a minimum at the somewhat higher value of 
$z=3.7$, however the minimum is very shallow, indicating a relative
insensitivity of the data to variations in $z$.  We conclude that both the
data for $R_x$ and $R_z$ are consistent with a dynamic exponent $z_{\rm loop}=3.4\pm 0.1$.

\begin{table}
  \centering 
  \begin{ruledtabular}
  \begin{tabular}{c|c|c|c|c}
$L_{\rm min}$ & $6$ & $8$ & $10$ &   $12$   \\\hline
$z_{\rm loop}$&$ 3.23\pm 0.05$ &$ 3.33\pm 0.07$ &$3.39\pm 0.11$&$3.38\pm 0.14$\\
\end{tabular}
\end{ruledtabular}
  \caption{Dynamic exponent $z_{\rm loop}$ from power law fits, $R_x\sim L^{-z}$,
   to system sizes $L=L_{\rm min} - 20$.}
  \label{t5}
\end{table}

\section{Discussion and Conclusions}

We have studied the equilibrium and dynamic critical behavior of the
zero magnetic field superconducting phase transition for a type-II superconductor with quenched
columnar disorder.  We have considered both the ``unscreened" XY model
in which $\lambda_0\to\infty$, and the ``strongly screened" loop model in which
$\lambda_0\sim\xi_0$.  A duality transformation establishes that these two models
are in the same {\it equilibrium} universality class.  Using numerical simulations
of the XY model, we find, in agreement with a generalized Harris criterion, that
the universality class of the  transition is different from the pure model,
and we find that scaling is anisotropic.  We find the value for the correlation length
exponent, $\nu=1.2\pm 0.1$, and for the anisotropy exponent, $\zeta=1.3\pm 0.1$.

Using the value of the critical temperature and the anisotropic scaling determined
from the equilibrium analysis, we carry out simulations at the critical point to 
determine the dynamic critical exponent $z$ of both the XY and loop models for local
Monte Carlo dynamic rules.  For the ``unscreened" XY model, with a single 
spin heat bath dynamics,
we find $z_{\rm XY}=2.6\pm 0.1$.  For the ``strongly screened" loop model, 
with a heat bath dynamics
applied to elementary loop excitations, we find $z_{\rm loop}=3.4\pm 0.1$.  

A similar random 3D XY model has been studied by Cha and Girvin\cite{Cha} 
in the context of the quantum phase transition in the two dimensional boson 
Hubbard model.  In their model disorder was introduced as uniformly distributed
random bonds in the $\hat z$ (imaginary time) direction, $J_{iz}$, so as to model 
bosons with random charging energy.  They found equilibrium critical exponents
 $\nu=1.0\pm 0.3$ and $\zeta=1.07\pm 0.03$ (our anisotropy exponent 
 $\zeta$ for the classical 3D
 model is equivalent to their ``quantum dynamic exponent" $z$ for the 2D quantum
 problem).  However their analysis for such a system with
anisotropic scaling, $\zeta>1$, was based on a more ad hoc approach of 
(i) trying various values of $\zeta$ and seeing which appeared to give the
best data collapse for systems of different size $L$, and (ii) measuring
real space correlations in a system of a fixed size and fitting to assumed power
law decays.  Their largest system size, $16^2\times 15$ is also smaller than ours
and they do not use the Wolff algorithm to accelerate their equilibration.
While it is possible that introducing the randomness differently (along $\hat z$
rather than in the $xy$ plane) might effect the universality class, we believe it
is more likely that this is not the case, and that our results are more systematic
and hence more accurate than those of Cha and Girvin.

Prokof'ev and Svistunov\cite{Svistunov} have simulated the loop model of Eq.(\ref{eHloop})
in the context of the same two dimensional disordered boson problem as
Cha and Girvin.  For their ``off-diagonal" disorder case they put
the disorder into the bonds along the $\hat z$ direction, making their
model dual to that of Cha and Girvin.  They report an anisotropy exponent
$\zeta=1.5\pm 0.2$, which agrees with ours within the estimated errors.  They
were unable to determine the correlation length exponent $\nu$.  We note that
while they use an accelerated ``worm" algorithm and have good statistics for
quite large system sizes, they determine their exponents by fitting to real
space correlation functions for their biggest size system, as did Cha and Girvin,
rather than doing any systematic finite size scaling that takes into account
the anisotropic scaling present in the model.

Experimental investigation of the zero field  transition with
colummnar disorder has been undertaken by K\"otzler and co-workers\cite{Kotzler}
for YBaCuO thin films.  Measuring the frequency dependent conductivity
transverse to the columnar disorder, which is expected to scale as\cite{Fisher}
$\sigma_\perp (\omega, T) = t^{(\zeta-z)\nu}\tilde\sigma(\omega t^{-z\nu})$
(where $t=(T-T_c)/T_c$),
they find\cite{note} the combinations $\nu\zeta=1.7$ and $z/\zeta=5.53$.
This compares with our values $\nu\zeta=1.56$, and $z/\zeta=2$ for the
unscreened XY model, and $z/\zeta=2.6$ for the strongly screened loop model.
Our value of $\nu\zeta$ is conceivably consistent with the experimental value, 
within possible errors.  However both of our values of $z/\zeta$ seem too
small.  It may be that our simple local Monte Carlo dynamics does not adequately
capture the true dynamics of a real superconductor.  On the other hand, if we
use our value of $\zeta=1.3$, then K\"otzler's results imply a dynamic exponent
of $z=7.2$, which seems extraordinarily large.

We may also compare our dynamic exponents with those obtained
from the disorder free model.  For the the strongly screened limit of the 
loop model, Lidmar {\it et al.}\cite{Wallin4} find the value $z_{\rm loop}\simeq 2.7$;
moreover they find this value to be insensitive to the presence of {\it uncorrelated}
point disorder.  For relaxational dynamics of the phase angle variable in the
XY model, a value of $z\approx 2$ is expected,\cite{Hohenberg} and this is
what was found in numerical simulations by Jensen {\it et al.}\cite{Minnhagen}
using a method similar to our  scaling of conductance, Eq.(\ref{eGmu}).
The result $z_{\rm loop}>z_{\rm XY}$ thus seems common for both the
pure and columnar disordered cases.

In our work we have considered only simple relaxational dynamics for the
phase angles of the unscreened XY model.  Two other possible dynamics might
be considered.  One would be to do a loop dynamics, similar to what we have
done here for the strongly screened loop model, only now as applied to the strongly
interacting loops of the unscreened XY model.  The other would be to use
resistively shunted junction (RSJ) dynamics for the phase angles of the XY model.  Both such
approaches have been previously used for the disorder free case.  
For both loop dynamics\cite{Wallin4,Weber}
and RSJ dynamics\cite{Minnhagen,Stroud} 
the dynamic exponent $z\simeq 1.5$ was found, smaller than
the value obtained by simple phase angle relaxational dynamics.  Investigating these
other dynamics for the case of columnar disorder remains for future work.  We 
only note here that if the above trend remains true for columnar disorder, and that 
these other dynamics reduce $z$ from that of relaxational dynamics, then it becomes
even harder to explain the large value of $z\zeta$ observed experimentally in
Ref.~\onlinecite{Kotzler}.

Finally, we note that similar {\it equilibrium} exponents to those found in this work
were also found for the case of an {\it unscreened} superconductor with columnar 
defects in a {\it finite} applied magnetic field.  For that case the values\cite{Wallin2} 
$\nu=1.0\pm0.1$ and $\zeta=1.25\pm0.1$ were found.  Although these are close
to the values we find here for {\it zero} applied field, there is no apparent reason that
the zero and finite field cases should be in the same universality class.  We also note
that once a finite field is applied, the duality between the unscreened and strongly
screened superconductor models, that exists for zero field, breaks down.

\section*{Acknowledgments}
We wish to thank U.\ C.\ T\"auber for originally suggesting this problem.
We thank T.\ J.\ Bullard, H.\ J.\ Jensen, U.\ C.\ T\"auber and M.\ Zamora for contributions 
at early stages of this work.
The work of A.\ V.\ and M.\ W.\ has been supported by the Swedish Research
Council, PDC, NSC, and the G{\"o}ran Gustafsson foundation.  S.\ T.\ 
acknowledges support from DOE grant DE-FG02-89ER14017, and travel support
from NSF INT-9901379.  H.\ W.\  acknowledges support from Swedish Research
Council contract 621-2001-2545.

\section*{Appendix A}

In this section we review the duality transformation\cite{Dasgupta,Kleinert,Savit} 
from ${\cal H}_{\rm XY}$ of Eq.(\ref{eHXY}) to ${\cal H}_{\rm loop}$ of Eq.(\ref{eHloop}).  Consider first a general $2\pi$ periodic interaction $V_{i\mu}(\phi)$ instead of the $-(J_{i\mu}/T)\cos(\phi)$ of Eq.(\ref{eHXY}).  For the generalized fixed twist boundary condition and the corresponding Hamiltonian of Eq.(\ref{eHXY2}), we can write the partition function as,
\begin{equation}
Z=\left(\prod_i\int_0^{2\pi}{d\theta^\prime_i\over 2\pi}\right) {\rm e}^{-\sum_{j\mu}V_{j\mu}(\theta^\prime_j-\theta^\prime_{j+\hat\mu}-\Delta_\mu/L_\mu)}\enspace.
\label{eZXY}
\end{equation}
where the $\theta^\prime_i$ obey periodic boundary conditions.
Defining the Fourier transform $\tilde V_{j\mu}$ by,
\begin{equation}
{\rm e}^{-V_{j\mu}(\phi)}\equiv \sum_{n_{j\mu}=-\infty}^{\infty}{\rm e}^{-\tilde V_{j\mu}(n_{j\mu})}{\rm e}^{in_{j\mu}\phi}\enspace,
\label{eFT}
\end{equation}
and substituting into Eq.(\ref{eZXY}) gives,
\begin{eqnarray}
Z&=&\sum_{\{n_{j\mu}\}}\left(\prod_{i}\int_0^{2\pi}{d\theta^\prime_i\over 2\pi}\right){\rm e}^{-\sum_{j\mu}\tilde V_{j\mu}(n_{j\mu})+i\sum_{j\mu}n_{j\mu}(\theta^\prime_j-\theta^\prime_{j+\hat\mu}-\Delta_\mu/L_\mu)}\\
&=&\sum_{\{n_{j\mu}\}}{\rm e}^{-\sum_{j\mu}\left[\tilde V_{j\mu}(n_{j\mu})+in_{j\mu}\Delta_\mu/L_\mu\right]}\left(\prod_{i}\int_0^{2\pi}{d\theta^\prime_i\over 2\pi}\right){\rm e}^{i\sum_{j\mu}n_{j\mu}(\theta^\prime_j-\theta^\prime_{j+\hat\mu})}\enspace.
\label{eZ2}
\end{eqnarray}
One is now free to do the integrals over the $\theta^\prime_j$.  The result is a product of Kronecker deltas constraining the variables $n_{j\mu}$ to be divergenceless, as in Eq.(\ref{en}).  Defining the ``winding numbers" $W_\mu$ by,
\begin{equation}
W_\mu \equiv {1\over L_\mu} \sum_i n_{i\mu}\enspace,
\label{eW}
\end{equation}
we get,
\begin{equation}
Z={\sum_{\{n_{j\mu}\}}}^\prime{\rm e}^{-\sum_{j\mu}\tilde V_{j\mu}(n_{j\mu})-i\sum_\mu W_\mu\Delta_\mu}\enspace,
\label{eZ3}
\end{equation}
where the prime on the summation denotes the divergenceless constraint of Eq.(\ref{en}).

A common choice for $V_{i\mu}(\phi)$ is the Villain interaction,\cite{Villain}
\begin{equation}
{\rm e}^{-V_{j\mu}(\phi)}\equiv \sum_{m=-\infty}^\infty{\rm e}^{-{J_{j\mu}\over 2T}(\phi-2\pi m)^2}\enspace.
\label{eV}
\end{equation}
In this case one has for its transform,
\begin{equation}
\tilde V_{i\mu}(n)={T\over 2 J_{i\mu}}n^2\enspace.
\label{eVT}
\end{equation}
The partition function of Eq,(\ref{eZ3}), with periodic boundary conditions
$\Delta_\mu=0$, then becomes,
\begin{equation}
Z={\sum_{\{n_{j\mu}\}}}^\prime {\rm e}^{-{1\over 2\tilde T}\sum_{j\mu}
g_{j\mu}n_{j\mu}^2}\enspace,
\label{eHloop2}
\end{equation}
with
\begin{equation}
g_{i\mu}/\tilde T = T/J_{i\mu}\enspace.
\label{egJ}
\end{equation}
The above is just a model of short ranged interacting loops with 
onsite repulsion $\sim n^2$ and inverted temperature scale $\tilde T\sim 1/T$.

For our simulatons, with $V_{i\mu}(\phi)=-(J_{i\mu}/T)\cos(\phi)$, one
has\cite{Savit}
\begin{equation}
{\rm e}^{-\tilde V_{i\mu}(n)}=I_n(J_{i\mu}/T)\enspace,
\label{eIn}
\end{equation}
where $I_n(x)$ is the modified Bessel function of the first kind.    Since $I_n(x)$
is an increasing function of $|n|$ for fixed $x$, the above similarly gives a short ranged loop
model with onsite repulsion.  It is this  interaction of
Eq.(\ref{eIn}) that we use in our dynamic simulations of the loop model in
Section IV.

We can now demonstrate several interesting results concerning phase coherence
in the XY model, by considering the behavior as a function of the twist $\Delta_\mu$.
The XY model is phase coherent when the total free energy ${\cal F}$ varies with $\Delta_\mu$.  Using ${\cal F}(\Delta_\mu)=-T\ln Z(\Delta_\mu)$, and Eq.(\ref{eZ3}) above, 
we find,
\begin{equation}
{1\over T}\left.{\partial{\cal F}\over\partial\Delta_\mu}\right|_{\Delta_\mu=0}=i\langle W_\mu\rangle_0\enspace,
\label{eF1}
\end{equation}
where $\langle\dots\rangle_0$ indicates an average in the ensemble with $\Delta_\mu=0$.
Now since $\partial{\cal F}/\partial\Delta_\mu$ must be a real quantity (as may be seen by considering its evaluation in the original XY model ${\cal H}_{\rm XY}$ of Eq.(\ref{eHXY2})), and since $\langle W_\mu\rangle_0$ must similarly be real (as may be seen by considering ${\cal H}_{\rm loop}$), the only way for Eq.(\ref{eF1}) to hold is if $\left.\partial{\cal F}/\partial\Delta_\mu\right|_{\Delta_\mu=0}=\langle W_\mu\rangle_0=0$.  This then demonstrates that $\Delta_\mu=0$, i.e. periodic boundary conditions on the $\theta_i$, is the twist that minimizes the free energy.

Finally, returning to Eq.(\ref{eZ3}), we note that in the {\it fluctuating twist} ensemble\cite{Olsson} for the XY model, in which $\Delta_\mu$ is averaged over as a thermally fluctuating degree of freedom, the corresponding loop model obeys the additional constraint of zero winding, $W_\mu=0$, in each individual configuration.  When viewing ${\cal H}_{\rm loop}$ as the Hamiltonian of vortex loops in a strongly screened superconductor, this corresponds to the ensemble in which the average internal magnetic field is constrained to vanish, $B_\mu=0$,
in each configuration.


\begin{thebibliography}{99}

\bibitem{Blatter}For reviews see, G.\ Blatter, M.\ V.\ Feigel'man,
  V.\ B.\ Geshkenbein, A.\ I.\ Larkin and V.\ M.\ Vinokur, Rev.
  Mod. Phys. {\bf 66}, 1125 (1994); E.\ H.\ Brandt, Rep. Prog. Phys.
  {\bf 58}, 1465 (1995); T.\ Nattermann and S.\ Scheidl, Adv. Phys.
 {\bf 49}, 607 (2000).

\bibitem{Nelson} D.\ R.\ Nelson and V.\ M.\ Vinokur, \prl {\bf 68},
2398 (1992); \prb {\bf 48}, 13060 (1993); {\em ibid} {\bf 61}, 5917
(2000).

\bibitem{Wallin1} J.\ Lidmar and M.\ Wallin, Europhys. Lett. {\bf 47}, 494 (1999).

\bibitem{Wallin2} A.\ Vestergren, J.\ Lidmar and M.\ Wallin, Phys. Rev. B {\bf 67}, 
 92501 (2003).

\bibitem{Wallin3} M.\ Wallin, E.\ S.\ S{\o}renson, S.\ M.\ Girvin and A.\ P.\ Young,
Phys. Rev. B {\bf 49}, 12115 (1994).

\bibitem{Chayes} J.\ T.\ Chayes, L.\ Chayes, D.\ S.\ Fisher and T.\
Spencer, \prl {\bf 57}, 2999 (1986); Commun.\ Math.\ Phys.\ {\bf 120}
501 (1989).

\bibitem{Harris} A.\ B.\ Harris, J. Phys. C {\bf 7}, 1671 (1974).

\bibitem{Dasgupta} C.\ Dasgupta and B.\ I.\ Halperin, Phys. Rev. Lett. {\bf 47}, 1556 (1981). 

\bibitem{Cha} M.-C.\ Cha and S.\ M.\ Girvin, \prb {\bf 49}, 9749 (1994).

\bibitem{Svistunov} N.\ Prokof'ev and B.\ Svistunov, Phys. Rev. Lett. {\bf 92}, 15703 (2004).

\bibitem{optical} D.\ Jaksch, C.\ Bruder,  J.\ I.\ Cirac, C.\ W.\ Gardiner and P.\ Zoller, 
Phys. Rev. Lett. {\bf 81}, 3108 (1998); B.\ Damski, J.\ Zakrzewski, L.\ Santos, P.\ Zoller 
and M.\ Lewenstein, Phys. Rev. Lett. {\bf 91}, 080403 (2003).

\bibitem{Sachdev} S.\ Sachdev, {\it Quantum Phase Transitions}, (Cambridge, 1999).

\bibitem{Teitel} Y.-H.\ Li and S.\ Teitel, Phys. Rev. Lett. {\bf 66}, 3301 (1991); 
  Phys. Rev. B {\bf 47}, 359 (1993); T.\ Chen and S.\ Teitel, Phys. Rev. B {\bf 55}, 15197 (1997). 
    
\bibitem{Fisher} D.\ S.\ Fisher, M.\ P.\ A.\ Fisher and D.\ A.\ Huse, Phys. Rev. B {\bf 43}, 
130 (1991).

\bibitem{Wallin4} J.\ Lidmar, M.\ Wallin, C.\ Wengel, S.\ M.\ Girvin and A.\ P.\ Young,
  Phys. Rev. B {\bf 58}, 2827 (1998).

\bibitem{Kleinert} H.\ Kleinert, {\it Gauge Fields in Condensed Matter}, (World
  Scientific, Singapore, 1989), Vol. 1; T.\ Banks, R.\ J.\ Myerson and J.\ Kogut, Nucl. 
  Phys. B {\bf 129}, 493 (1977).
  
\bibitem{Savit} R.\ Savit, Phys. Rev. B {\bf 17}, 1340 (1978).

\bibitem{Villain} J.\ Villain, J. Phys. Paris {\bf 36}, 581 (1975).

\bibitem{Young} H.\ Rieger and A.\ P.\ Young, Phys. Rev. Lett. {\bf 72}, 4141 (1994);
  M.\ Guo, R.\ N.\ Bhatt and D.\ A.\ Huse, Phys. Rev. Lett. {\bf 72}, 4137 (1994).

\bibitem{Nightingale} M.\ P.\ Nightingale and H.\ W.\ Bl\"ote, Phys. Rev. Lett. {\bf 60},
  1562 (1988).

\bibitem{Binder} K.\ Binder, Phys. Rev. Lett. {\bf 47}, 693 (1981).

\bibitem{Olson} T.\ Olson and A.\ P.\ Young, \prb {\bf 61}, 12467 (2000).

\bibitem{Reger} J.\ D.\ Reger, T.\ A.\ Tokuyasu, A.\ P.\ Young and M.\ P.\ A.\ Fisher, Phys. 
   Rev. B {\bf 44}, 7147 (1991).

\bibitem{Wolff} U.\ Wolff, \prl {\bf 62}, 361 (1989).


\bibitem{Huse} D.\ A.\ Huse, private communication.

\bibitem{Hohenberg} P.\ C.\ Hohenberg and B.\ I.\ Halperin, Rev. Mod. Phys. {\bf 49}, 435
  (1977).

\bibitem{Minnhagen}  L.\ M.\ Jensen, B.\ J.\ Kim and P.\ Minnhagen, Phys. Rev. B {\bf 61},
  15412 (2000) and Europhys. Lett. {\bf 49}, 644 (2000).
  
\bibitem{Minnhagen2} P.\ Minnhagen, B.\ J.\ Kim and H.\ Weber, Phys. Rev. Lett. {\bf 87},
  37002 (2001).

\bibitem{Young2} A.\ P.\ Young, in {\it Random Magnetism, High-Temperature
Superconductivity"}, eds. W.\ P.\ Beyerman, N.\ L.\ Huang-Liu and
D.\ E.\ MacLaughlin (World Scientific, Singapore, 1994).

\bibitem{Kotzler} G.\ Nakielski, A.\ Rickertsen, T.\ Steinborn, J.\
Wiesner, G.\ Wirth, A.\ G.\ M.\ Jansen and J.\ K\"otzler, Phys.\ Rev.\
Lett.\ {\bf 76}, 2567 (1996); J.\ K\"otzler, in {\it Advances in Solid State Physics} 39,
ed. B.\ Kramer, (F.\ Vieweg \& Sohn Verlags-GmbH, Braunschweig/Wiesbaden, 1999),
p.\ 371  (also as cond-mat/9904279). 

\bibitem{note} Due to differences of notation, the exponents $\nu$ and $z$ in Ref. \protect\onlinecite{Kotzler} correspond to our $\nu\zeta$, and $z/\zeta$.

\bibitem{Weber} H.\ Weber and H.\ J.\ Jensen, Phys. Rev. Lett. {\bf 78}, 2620 (1997).

\bibitem{Stroud} K.\ H.\ Lee and D.\ Stroud, Phys. Rev. B {\bf 46}, 5699 (1992).

\bibitem{Olsson} P.\ Olsson, Phys. Rev. Lett. {\bf 73}, 3339 (1994) and Phys.
  Rev. B {\bf 52}, 4511 (1995).

\end{thebibliography}
\end{document}